# General-purpose foundation models surpass specialized ones on chemical property prediction tasks

Stepan N. Ostarkov[1]*, Gregorii A. Semakin[1], Ivan A. Bespalov[1], Michael G. Medvedev[1]*
[1]N.D. Zelinsky Institute of Organic Chemistry, Russian Academy of Science, 119334, Moscow, Russia

**Abstract**

Foundation models (FMs) learn reusable representations from large collections of unlabeled data. In chemistry the corresponding data mainly consists of molecular structures, which are abundant, whereas experimental measurements are scarce. Chemical FMs such as ChemBERTa[1] and MoLFormer[2] were therefore trained on diverse molecular structures to enable fine-tuning on small experimental datasets (<100 datapoints in most realistic scenarios) and subsequent property prediction. At the same time, general-purpose FMs such as GPT-6 Astra and Claude Opus 5.5 were trained on virtually all publicly available data, including humankind's knowledge on molecular properties and laws governing them. Here we compare accuracies of chemical and general-purpose FMs on tasks from various areas of chemistry. Strikingly, general-purpose GPT-6 Astra outperforms the currently eligible chemical FMs on five physicochemical and quantum-chemical regression tasks when only 60 experimental datapoints are available. Only when provided 10x datapoints do the best chemical FMs reach performance of GPT-6 Astra: on 1,000 and 10,000 points they show performance on par with Astra on 60 and 1,000 points, respectively. Thus, counterintuitively, frontier general-purpose FMs are better suited for guiding chemical research in data-scarce areas than specialized chemical FMs, suggesting that learning relevant primitives (e.g. chemical structures) does not substitute training on relevant humankind knowledge.

Foundation models (FMs) acquire knowledge and capabilities through large-scale training and transfer them to a range of downstream tasks[3]. For current large language models (LLMs), the training pipeline extends well beyond next-token prediction: it combines broad pretraining with instruction-following examples, human feedback, mathematical reasoning tasks, and training for tool use and agentic behavior.[4,5] Although these stages target specific capabilities, they may also give rise to emergent abilities, such as reasoning about molecular geometry and molecular interactions.[6,7]

The practical promise of an FM is that this prior training reduces the amount of task-specific data needed for a new prediction problem. Instead of learning everything from a small labelled dataset, a downstream predictor can draw on knowledge acquired during pretraining to make predictions when experimental data are limited.[8] Thus, the value of an FM is determined by what additional information this training preserved for reuse.

In chemistry experimental data is particularly scarce, with most research decisions being based on <100 datapoints, which might well take several years to collect. Moreover, a dataset measured for one property does not automatically provide labels for another. Predicting solubility, enzyme inhibition, or electronic energies requires different training datasets. These factors make reuse of prior chemical knowledge attractive, especially when only a small set of compounds has been characterized.[9,10]

This is where chemical FMs come into play. They have been extensively used across medicinal, environmental and materials chemistry, with ChemBERTa and MolLFormer original articles currently having 1700 and 1022 citations according to Google Scholar. In environmental toxicology, MoLFormer-XL and ChemBERTa-2 have been adapted to model PFAS bioactivity[11]. In medicinal chemistry, MoLFormer helped select compounds for experimental screening against *S. pneumoniae*[12], as well as *S. aureus* and *E. coli*[13]. Molecular representations also support formulation research: GROVER embeddings have been used to represent lipid components when predicting the transfection efficiency of mRNA-loaded lipid nanoparticles.[14]

[1] Corresponding author(-s): ostarkovstepan@gmail.com, medvedev.m.g@gmail.com.

Training of chemical FMs is based on the fact that unlike experimental data, molecular structures can be generated at scale, so they are usually pretrained on objectives to reconstruct masked molecular tokens or graph attributes, or predict simple descriptors readily calculated from SMILES[1,15,16]. Although these objectives expose models to molecular structures and easily computed properties, downstream tasks often require predictions that go beyond these pretraining targets. Chemical behavior depends on relationships between structure, interactions and physical conditions, quantitative account for which requires chemical understanding, mathematics, and reasoning. Learning the regularities of molecular structures may contribute to this understanding and suggest areas, where chemical FMs can be expected to excel: studies employing them frequently target biochemistry-related classification tasks and often include hundreds or thousands of data points: a setting rarely encountered in fresh areas of chemistry, where data is scarce and the target property is usually a number, not a label. Whether molecular pretraining supplies this broader foundation for regression tasks in areas other than biochemistry remains an open question.

Here, we compare frontier chemical FMs (see the list below) with frontier general-purpose FMs (GPT-6 Astra, Claude Opus 5.5, and Gemini 4 Argon) and classic human-crafted molecular representations, which were used long before the concept of FMs emerged.

The models were tested on six prediction tasks based on published and widely used benchmarks, with first five covering regression properties: (1) Lipophilicity index which is a physicochemical property very important in drug design, (2) ESOL benchmark for solubility in water, (3) FreeSolv benchmark for free energy of hydration, (4) energies of lowest unoccupied molecular orbital ($\varepsilon_{LUMO}$) from QM9, and (5) internal energies of molecules at 0K ($U_0$) from QM9, and (6) BACE benchmark covering biochemical classification property – BACE1 protein inhibition – selected as a control where chemical FMs are expected to work well. To simulate the amount of data available in a realistic chemical project, we extracted 100 random datapoints (molecules and their properties) from each dataset and split them into training, validation, and test sets in proportion of 5:1:4. Thus, we consider that 60 experimental values are known (available for training the model) for each property. $R^2$ is used as a measure of performance on regression benchmarks (1-5), while for the only classification benchmark (BACE) we use 2×(ROC-AUC − 0.5) which, like $R^2$, equals 1 for perfect prediction, while 0 for both metrics correspond to predicting an average (0.5 positive class probability for classification or mean value for regression)

In frozen setting, each chemical FM (GIN[15], GROVER[17], MoLFormer, MoLFormer XL[2], MoLFormer-c3[18], SELFormer[19], CLAMP[20], ChemFM-1B[21], ChemBERTa-2[1], ChemBERTa-3[18], CheMeleon[16]) or human-crafted representation (RDKit descriptors[22] and Morgan fingerprints[23]) was used to compute molecular embeddings (vector representations) for all the molecules selected from each benchmark. Then prediction heads were trained and validated on the corresponding subsets (see Methods) and used to predict the corresponding property on the test subset. The following heads were considered: linear, MLP, XGBoost, Gaussian process with RBF or Matérn kernel, and two tabular FMs pretrained for tabular prediction: TabICLv2[24] and TabPFN-3.5[25]. To assess the quality of molecular representations emerged in general-purpose FMs, we also included Gemini Embedding 2, which was treated identically. Further we evaluated full fine-tuning[26] and low-rank adaptation (LoRA[27]) for all chemical FMs, using their own prediction heads.

General-purpose FMs were tested in a few-shot regime: they were prompted with a list of SMILES and experimental properties for all compounds in the train and validation subsets and asked to predict properties for compounds in the test subsets (also supplied as SMILES). Each benchmark was prompted separately. Detailed procedures and prompts are available in Methods and SI.

Figure 1 shows distributions of test scores for (1) the best (according to validation scores) combination of human-crafted (classic) baseline with a head which is not an FM – grey, (2) the best combination of a classic baseline with an FM head (TabICLv2 or TabPFN-3.5) – black, (3) the best combination of chemical FM, its adaptation and prediction head – blue, (4) the best combination of Gemini Embedding 2 and prediction head – pale purple, and (5-8) the three general-purpose foundation models – shades of purple. Light grey distributions show performance of general-purpose FMs under negative control when correct answers for train and validation molecules were randomly shuffled: good performance under these conditions would suggest that a model used information beyond the supplied labels, which may include prior target knowledge. All distributions are derived from 15 runs with identical parameters using differing splits between subsets. Source data for Figure 1 is available in SI.

Strikingly, GPT-6 Astra shows predictive performance higher than the best chemical FM on all test sets except for BACE (Fig. 1). Their performances are, respectively, 0.99 vs. 0.67 on total energy $U_0$, 0.85 vs. 0.53 on $\varepsilon_{LUMO}$, 0.97 vs. 0.63 on FreeSolv, 0.92 vs. -4.66 on ESOL, 0.60 vs. 0.28 on Lipophilicity, and 0.47 vs. 0.70 on BACE1 inhibition. Gemini 4 Argon mirrors behavior of GPT-6 Astra, but in most cases undershoots it. Claude Opus 5.5 also showed a very good performance (even overshooting GPT-6 Astra on ESOL and FreeSolv), but its good predictions under negative control (with shuffled data) suggest that it might have memorized correct answers for these tests during its training; it also refused to answer some prompts regarding ESOL, Lipophilicity and BACE. These results suggest that a general-purpose LLM can serve as a strong starting point for these chemical prediction tasks, enabling high performance without training a new chemistry-specific model.

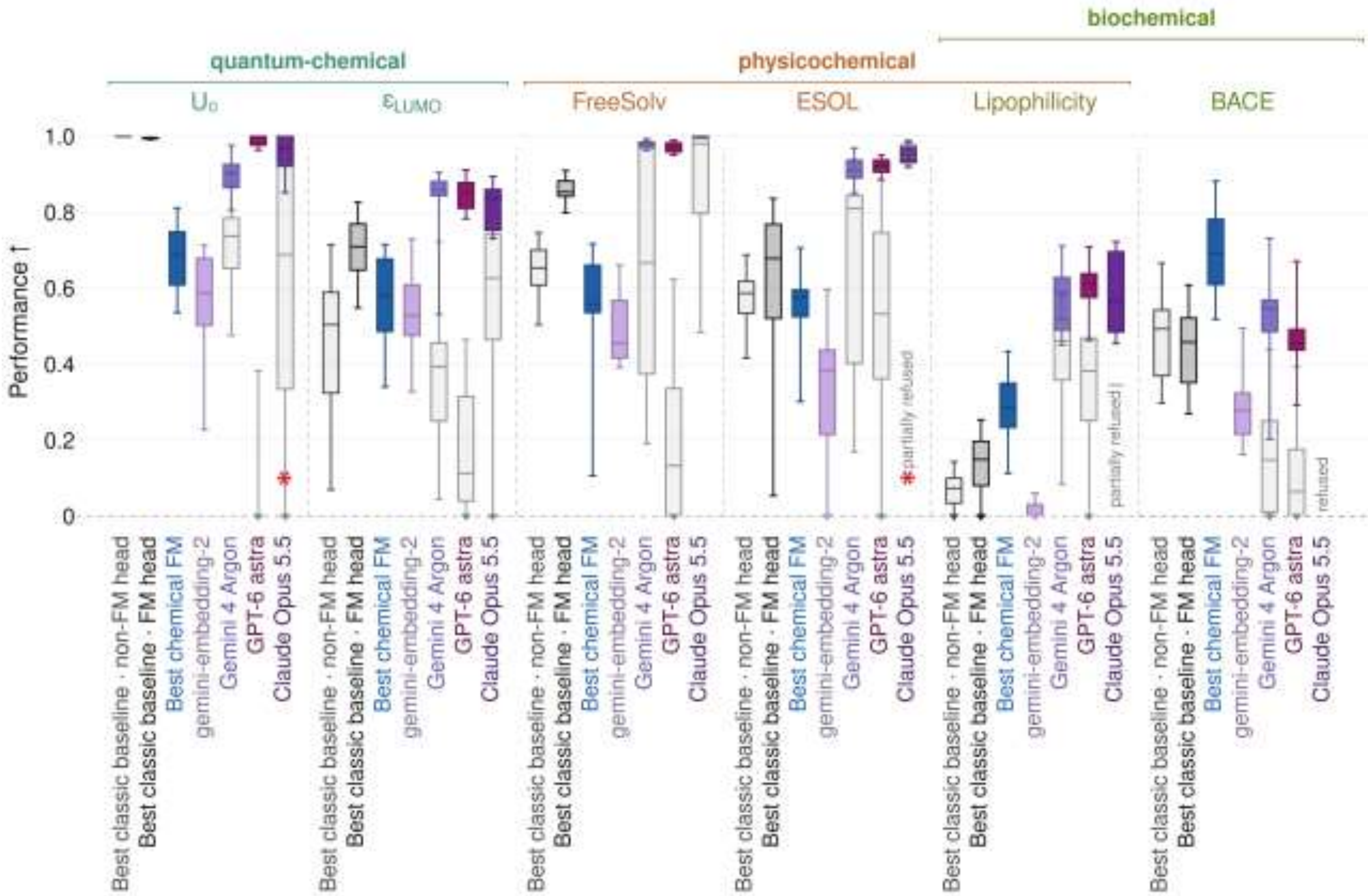

***Fig. 1 | Molecular property prediction from 60 labelled datapoints**. Scores are $R^2$ for regression and 2 × ROC-AUC − 1 for classification (BACE); dashed lines mark randomly guessing the class for classification or predicting the mean for regression. Boxes show the median and interquartile range; whiskers span the observed scores. Individual point markers are omitted. Grey boxes show performance after shuffling the labels of labelled datapoints. Downward triangles denote points below y=0, see fig S1 for full data. Red asterisks (*) denote suspected memorization (see Methods for details). 'Refused' indicates that the model declined to predict a target property in all fifteen attempts; 'Partially refused' indicates that the model declined to predict a target property in some attempts, see SI for details.*

However, chemical FMs retain an advantage on BACE in the current comparison. ChemBERTa-2 paired with a TabPFN head reaches mean ROC-AUC of 0.85, compared with 0.74 for Astra, and leads in all 15 repeats. This advantage may be largely attributed to the fact that datasets frequently used for chemical foundation model training (PubChem, ZINC, ChEMBL) usually consist of biochemically relevant structures.

Unexpectedly high performance of all general-purpose FMs on shuffled labels (negative control) deserves a separate discussion. Frontier models, such as Opus 5.5., supplied with a nonsensical input, may decide to ignore it, treating it as misaligned with their internal world model and instead proceed to reason straight from the structure. Indeed, a chemist supplied with examples of highly polar molecules labeled as hydrophobic and *vice versa* would still remember that usually highly polar molecules are hydrophilic. Thus, the observed high performances do not necessary imply memorization.

The general-purpose embedding models provide a complementary route to prediction. Since embeddings from proprietary frontier LLMs, such as Opus and Astra, cannot be obtained, we selected a dedicated embedding model, gemini-embedding-2 for comparison. Its performance is quite low, sometimes near-random, and it appears on par with chemical FMs only for $\varepsilon_{LUMO}$. This finding highlights the importance of reasoning performed by general-purpose FMs and their reliance on prior knowledge in data-scarce settings.

Pretraining of the prediction head also affects their performance. For the same RDKit descriptors, using TabPFN or TabICL as a prediction head raises mean $R^2$ from 0.57 to 0.64 for ESOL, from 0.65 to 0.86 for FreeSolv, and from 0.52 to 0.70 for $\varepsilon_{LUMO}$, suggesting tabular FMs as highly-valuable instruments for few-shot molecular predictions. Tabular models used a variety of available tabular data in the pretraining stage. Such data could have included chemical or chemically-relevant tables, allowing these FMs to preserve valuable chemical knowledge, just as general-purpose FMs do.

As discussed above, the number of relevant experimental data for any particular chemical task rarely exceeds a hundred. There are, however, some better explored chemical tasks, for which larger datasets exist: the six benchmarks we have extracted our test sets from here provide examples of such. Moreover, novel experimental approaches including self-driving labs[28] and microfluidics[29,30] potentially allow scientists to collect data at previously unprecedented scale. It is therefore interesting to analyze how performance of the methods discussed above changes when the number of datapoints grows.

We therefore examined (Fig. 2) how additional labelled data affects the models' performance for $\varepsilon_{LUMO}$, for which no model reaches near-exact accuracy at 60 datapoints (Fig. 1) and there are plenty of datapoints available in QM9. Chemical FMs were trained and validated with frozen embeddings and a prediction head, full fine-tuning, and low-rank adaptation, on 100, 1,000, and 10,000 datapoints (with 9:1 train-validation split) and then tested on 25, 250, or 2,500 datapoints, respectively. Models based on RDKit descriptors and Gemini Embeddings 2 were trained and tested on the same subsets. GPT-6 Astra was tested on predicting $\varepsilon_{LUMO}$ of 250 molecules when provided 1,000 labelled ones.

With 100 labeled molecules provided, the strongest chemical FM result reaches $R^2$ 0.72, below GPT-6 Astra supplied with 60 labels, which shows performance of 0.85. Notably, in this setting both full fine tuning and LoRA provide worse results than prediction heads on top op a frozen chemical FM. At 1000 labelled datapoints GPT-6 Astra is still the best model, but some chemical FMs come very close to it: 0.93 vs. 0.91 for SELFormer with LoRA. Strikingly, this performance of SELFormer provided with 1,000 datapoints is only 0.06 above performance of GPT-6 Astra provided with only 60 datapoints. Upon proceeding to 10,000 labelled molecules, accuracy of chemical FMs continue to rise and many of them converge above 0.9: ChemFM-1B with LoRA reaches 0.97. LoRA generally helps chemical FMs more than full fine-tuning in this study; the latter sometimes deteriorates the model, particularly when data count is low (see GIN, CLAMP, and ChemBERTa-2) or may lead to divergence during training, as it was observed for ChemFM-1B. When transitioning from 60 to 100 data points, some models (MoLFormer-c3 in full fine-tune regime, as well as ChemBERTa-2 and CLAMP with LoRA) exhibit a performance drop, highlighting sensitivity of FM adaptation to data shift.

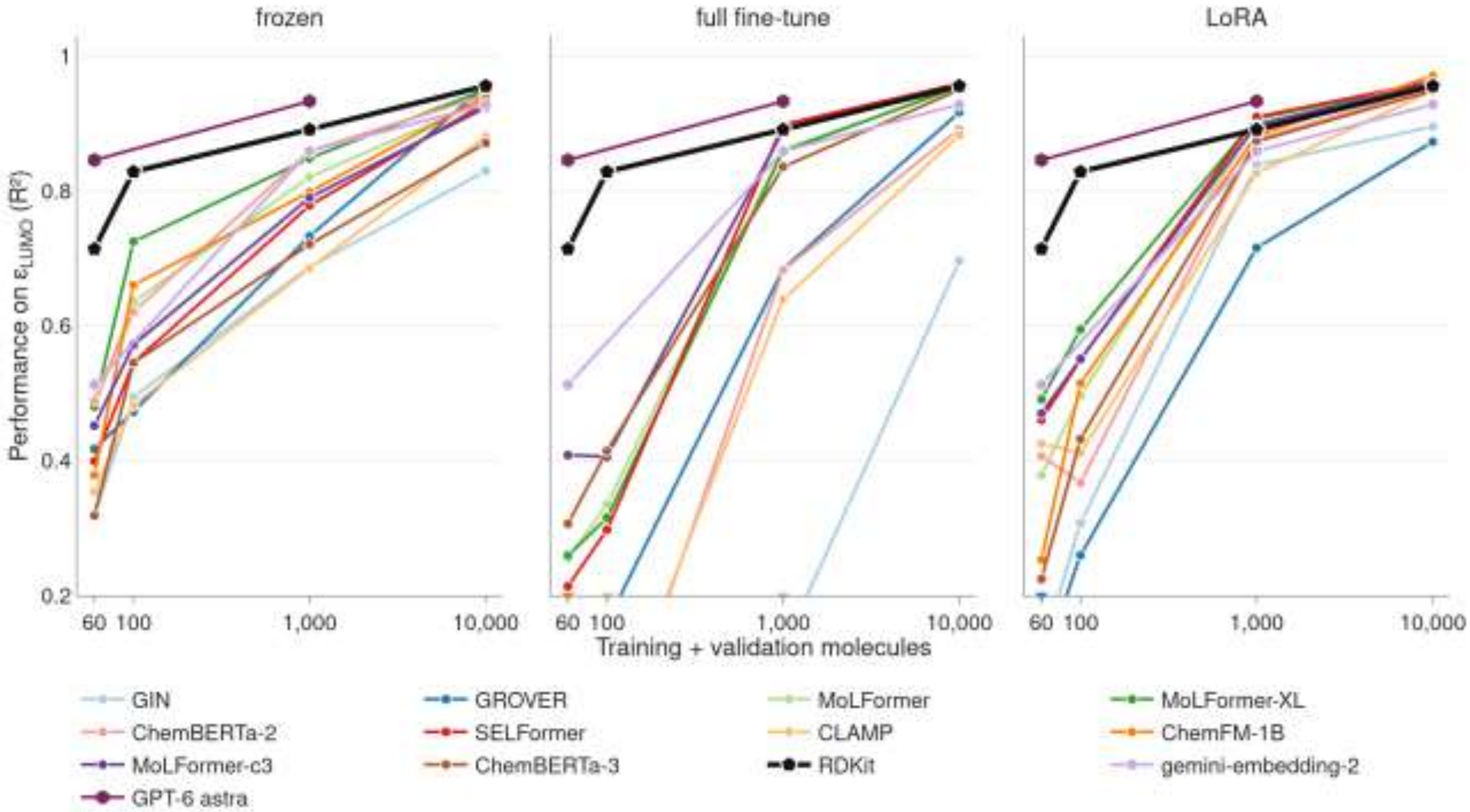

**Fig. 2 | Substantial amounts of additional labelled data can improve chemical foundation models performance for $\varepsilon_{LUMO}$ property prediction.** Mean $R^2$ across five repeats is shown for frozen representations (left), full fine-tuning (middle) and low-rank adaptation (LoRA, right), at 60, 100, 1000 and 10000 molecules in training and validation subsets together. Black lines show RDKit descriptors with the best-performing prediction head. GPT-6 Astra's performance corresponds to few-shot prediction and is identical in all three plots. Downward triangles mark values below the plotted $R^2$ range; Supplementary Fig. 2 shows the same results without clipping. ChemFM-1B diverged to NaN in all our full-fine-tuning attempts, except for 60 labelled datapoints.

Notably, as a dataset size continues to rise, few-shot learning becomes less feasible, since model reasoning scales accordingly and starting from some dataset scale it may not fit within the fixed output-length cap. A fitted molecular predictor, in its turn, can evaluate queries in batches. Together with the rise of chemical FMs performance for larger datasets, this opens a niche where they can excel.

For the more common data-scarce regression regime, however, general-purpose FMs and even human-crafted descriptors show higher performance. Supply a general-purpose FM, such as GPT-6 Astra, with the property name, units, all the properties that you deem reasonable, along with properly chosen molecular examples, and evaluate on held-out query molecules. These steps describe the workflow tested here and do not require a new pretraining campaign.

In a more general sense, our results suggest that the much broader training of general-purpose FMs allowed them to become better in the low-data regime, than much more specialized chemical FMs. While the latter were trained primarily on molecular structures, training corpuses of the former contained much more declarative chemical knowledge, including various properties and known scientific laws. Our results suggest that knowing which objects exist in a domain, while definitely not useless, does not substitute knowing their properties and relations between them. Training only on these objects (e.g. molecules) may be compared to training an LLM on the list of words without any sentences. Thus, a new paradigm for training specialized chemical FMs is required; meanwhile, general-purpose FMs remain the best option when data is scarce.

# Data availability

The benchmark datasets are available from the sources cited in Methods. Source data for the figures is available at https://github.com/TheorChemGroup/FewShot4Chem

# Code availability

Code for model evaluation and figure generation is available at https://github.com/TheorChemGroup/FewShot4Chem

# Acknowledgements

The authors are grateful to the Russian Science Foundation for financial support (grant #22-73-10124-P). This research was supported in part through computational resources of HPC facilities at NRU HSE.

# Author contributions

O.S.N. performed all the experimental work and wrote the article text. M.M.G. provided scientific guidance for the project. S.G.A. helped to polish and reviewed the article text. B. I. A. contributed to the original idea and provided additional guidance.

# Declaration of competing interests

The authors declare no competing interests.

# AI use statement

While writing the code and the article text, authors used several AI-based instruments, such as GPT-6 Astra, GPT-6.1-Sol via Codex, Claude Sonnet 4.5, 4.5 and 5.0, Claude Opus 4.6, 4.7, 4.8, 5.0 via Claude Code. Nevertheless, all authors take complete responsibility for all the results and claims in the article. The final version of the article was reviewed and polished by the authors.

# Methods

The main comparison comprised BACE inhibition classification[31], Lipophilicity[10], ESOL[32], FreeSolv[33], and $\varepsilon_{LUMO}$ and $U_0$ from QM9[34]. 100 molecules were sampled randomly from each dataset, then 15 splits of 50 training, 10 validation and 40 test molecules were obtained from them.

The eleven-model chemical panel comprised ChemBERTa-2-MTR[1], ChemBERTa-3-MLM and MoLFormer-c3[18], MoLFormer-base and MoLFormer-XL[2], ChemFM-1B[21], SELFormer[19], GROVER[17], GIN-pretrained[15], CLAMP[20], and CheMeleon[16]. We evaluated frozen representations, full fine-tuning and low-rank adaptation[27]; configurations were selected using comparable validation scores across fifteen repeats. We also tested 43 RDKit descriptors and 2,048-bit radius-two Morgan fingerprints. GROVER used no appended descriptor vector.

Further details on dataset, checkpoints, splits, and prompts are provided in the Supplementary Methods and in repository: https://github.com/TheorChemGroup/FewShot4Chem.

We compared regularized linear predictors, a multilayer perceptron, gradient boosting[35], Gaussian processes with RBF and Matérn kernels[36], TabPFN[37,38] and TabICL[24,39]. FM heads correspond to TabICLv2 and TabPFN-3.5.

Both regression and BACE classification were evaluated with these seven methods. Feature standardization was fitted on the training and validation molecules without using validation labels; supervised heads were fitted on the training split. Each frozen prediction head, and adaptation strategy (none, full fine-tune, LoRA) was selected by its mean validation score across fifteen splits, and its test scores were reported.

The three general-purpose language models were GPT-6 Astra, Gemini 4 Argon and Claude Opus 5.5, with service identifiers openai/gpt-6-astra, google/gemini-4-argon and anthropic/claude-opus-5.5. Each prompt supplied 60 labelled examples and 40 query molecules, using normal or shuffled support labels. To form a prompt, training and validation molecules were combined as labelled support. Prompts contained canonical SMILES with support labels, the property name and units, and unlabelled query compounds; benchmark dataset names were withheld. Support labels were rounded for presentation, while evaluation used unrounded query labels as targets. Only complete prediction vectors with matching query identifiers were scored; incomplete, misidentified, refused and truncated responses were excluded.

On Fig. 1, red asterisks mark overlap between the normal-label and shuffled-label two-sided 95% Student-t confidence intervals of mean repeat-level test performance, calculated separately from the available scored repeats in each mode. Intervals are calculated across scored split-level repeats and are not estimated with fewer than two repeats.

For the $\varepsilon_{LUMO}$ size study, 12500 molecules were randomly sampled from QM9. From them, the samples of the size 125 and 1250 were randomly drawn. Sets of all sizes (125, 1250, and 12500) were divided into the training, validation, and test sets in 9:1:2.5 proportion. These sets were then employed to train the models following the same protocol as in the main comparison.

# Supplementary Information

General-purpose foundation models surpass specialized ones on chemical property prediction tasks

## S1. Datasets and repeated splits

The main comparison uses six targets from five molecular datasets. One cohort of 100 molecules is fixed for each dataset. The two QM9 targets share the same molecular cohort and splits. Cohorts were sampled without replacement with NumPy default_rng seed 2032025 and retained in source-dataset order. For the QM9 cohort, 1,250 molecules were first drawn using seed 2032025, then 100 molecules were drawn from this subset using seed 2032025.

Each of the fifteen repeats (s=0…14) divides that cohort into 50 fitting, 10 validation and 40 test molecules with the seed 2032025+s. The language models receive the combined 60 fitting-plus-validation labels. Dataset sources are cited in the main text [9,24–27].

Targets and their units are described in the table S1 below.

**Table S1. Targets and units**

| Dataset | Target | Units or labels | Source pool for cohort sampling |
|---|---|---|---|
| BACE | BACE1 inhibition | 0=non-inhibitor; 1=inhibitor | 1513 |
| Lipophilicity | Octanol/water logD at pH 7.4 | base-10 logarithm | 4200 |
| ESOL | Aqueous logS | log10 of solubility in mol/L | 1128 |
| FreeSolv | Hydration free energy | kcal/mol | 642 |
| QM9 | Lowest unoccupied molecular orbital energy, $\varepsilon_{LUMO}$ | eV | 1250 |
| QM9 | Internal energy at 0 K, $U_0$ | eV | 1250 |

## S2. Molecular representations and checkpoints

Frozen representations are calculated once for the fixed molecular cohorts and reused across splits and downstream heads. The 11 chemical models are compared with 43 RDKit descriptors, radius-two 2, 048-bit Morgan fingerprints and a 3, 072-dimensional general-purpose embedding. GROVER receives no appended RDKit descriptor vector. CheMeleon uses the released message-passing weights with mean aggregation to obtain embeddings. The identifiers in Table S2 describe the recorded configurations; they do not establish that every older experiment used an independently verified weight digest.

**Table S2. Representation identifiers and output dimensions**

| Representation | Checkpoint or definition | Dimension |
|---|---|---|
| ChemBERTa-2 [1] | DeepChem/ChemBERTa-77M-MTR | 384 |
| ChemBERTa-3 [13] | DeepChem/ChemBERTa-100M-MLM | 768 |
| ChemFM-1B [16] | ChemFM/ChemFM-1B | 2048 |
| CLAMP [15] | Compound encoder: checkpoint.pt and hp.json; SHA-256 digests in the accompanying checkpoint manifest. | 768 |
| GIN pretrained [10] | supervised_contextpred.pth | 300 |
| GROVER-base [12] | model_base.pt | 4096 |

| Representation | Checkpoint or definition | Dimension |
|---|---|---|
| MoLFormer-c3 [13] | DeepChem/MoLFormer-c3-1.1B; revision 9f1b9ea3590833bd0ea1a70e789c5d3da11ba7ed; delegated code 7b12d946c181a37f6012b9dc3b002275de070314 | 768 |
| MoLFormer-XL [2] | ibm/MoLFormer-XL-both-10pct; revision 7b12d946c181a37f6012b9dc3b002275de070314 | 768 |
| MoLFormer original [2] | N-Step-Checkpoint_3_30000.ckpt; original vocabulary | 768 |
| SELFormer [14] | HUBioDataLab/SELFormer | 768 |
| RDKit descriptors [17] | RDKit Properties descriptor set | 43 |
| Morgan fingerprints [18] | radius 2, 2048 bits | 2048 |
| CheMeleon [11] | chemeleon_mp.pt; Zenodo record 15460715 | 2048 |
| Gemini Embedding 2 | google/gemini-embedding-2 | 3072 |

## S3. Prediction heads and preprocessing

Each frozen representation is evaluated with seven heads. Feature standardization uses means and standard deviations fitted on the 60 fitting-plus-validation feature rows; test features are excluded. Supervised fitting uses the 50 fitting labels, while the 10 validation labels are used for early stopping, hyperparameter and model selection. A head's own internal preprocessing is retained. For seeded heads, the repeat seed is 2032025+s.

**Table S3. Head configurations in the 100-molecule comparison**

| Head | Recorded configuration |
|---|---|
| Linear | RidgeCV for regression; LogisticRegressionCV for BACE. Fivefold internal CV. Candidate α or C: 0.001, 0.01, 0.1, 1, 10, 100; classification max_iter 2000. Training loss: L2-penalized squared error (regression) or L2-penalized logistic loss (BACE). |
| MLP | Two hidden layers of 256 units, GELU, dropout 0.1; AdamW, learning rate 0.001, weight decay 0.0001, batch 64; up to 200 epochs, validation-loss early stopping with patience 10. Training loss: mean squared error (regression) or binary cross-entropy with logits (BACE). Last epoch weights are used. |
| XGBoost [28] | 100 boosting rounds; depth 3, eta 0.3, subsample 0.7, $L_2$ penalty 0.01, hist tree method; reg:squarederror or binary:logistic. Training objectives: squared error (regression) and binary logistic loss (BACE). |
| GP-RBF [29] | Sparse variational GP; up to 1024 inducing points, capped by sample count; 200 optimization iterations, learning rate 0.01, batch 1024. Training loss: negative variational ELBO with Gaussian likelihood (regression, with standardized fitting targets) or Bernoulli likelihood (BACE). |
| GP-Matérn [29] | Same optimization settings, with $\nu$=2.5. The same negative variational ELBO and likelihoods are used. |
| TabPFN [20] | TabPFN-3.5; package 9.2.0; tabpfn-v3.5-20260909.safetensors; ignore_pretraining_limits=True. |

| Head | Recorded configuration |
|---|---|
| TabICL [19] | TabICLv2; package 2.2.0; tabicl-regressor-v2-20260212.ckpt or tabicl-classifier-v2-20260212.ckpt. |

## S4. Adaptation

To perform full fine-tuning, a 256→256→1 MLP with GeLU activation was added to the model, except for CheMeleon, the model that was specifically designed for fast fine-tune, which follows its native message-passing implementation with a fresh 256-unit feed-forward head.

Ranks 4, 8 and 16 are chosen for LoRA, the validation metric is used to choose the best.

Both full fine-tune and LoRA used AdamW optimizer with learning rate $10^{-5}$ and weight decay 0.01, MSE loss for regression and cross entropy for classification. Early stopping with patience 8 epochs was employed. The epoch was chosen by best validation metric.

## S5. Few-shot molecular tables and exact prompts

The following prompt template was used for property prediction:

```
Below is a table of 60 molecules, written as SMILES, each with a known value of one property:
  [Property description]

Use this table as training data to predict the same property for the 40 query molecules listed after it.
[Classification/regression request] Predict every query molecule, even when uncertain. Do not write
or run code; answer directly.

TRAINING TABLE (SMILES<TAB>value)
[60 saved support rows]

QUERY MOLECULES (id<TAB>SMILES)
[40 saved query rows]

Return JSON of the form {"predictions": [{"id": "<query id>", "value": <number>}, ...]} with
exactly one entry per query id.
```

Property descriptions are provided in table S5 below.

**Table S4. Property descriptions used in the prompts**

| Property | Verbatim property description |
|---|---|
| active | inhibition of human beta-secretase 1 (BACE-1): 1 = inhibitor, 0 = non-inhibitor |
| logD | logD: octanol/water distribution coefficient at pH 7.4 (base-10 logarithm) |
| logS | logS: the base-10 logarithm of aqueous solubility, with solubility in mol/L |
| dG_hydr | hydration free energy, in kcal/mol |
| epsilon_lumo | LUMO energy, in eV |
| U_0 | internal energy at 0 K, in eV |

For classification, we used the following request:

The values are 1 or 0. For each query molecule, give the PROBABILITY (a number from 0 to 1) that its value is 1.

And for regression:

For each query molecule, give a numeric prediction of the property in the same units. Predict every query molecule, even when uncertain.

Training and validation sets were combined to form support set. Support and query examples were shuffled (preserving the labels and support/query identity) using Python random.Random(1000+s). For each target and split, the normal and shuffled prompts have the same structures, query identifiers and unrounded query labels. The shuffled prompt permutes only the support labels using random.Random(90000+s), preserving their multiset.

Binary labels are written as integers. For continuous labels the number of decimals is min(4, max(2, 4−floor(log10 R))), where R denotes the range between the largest and the smallest label value. Scoring retains the unrounded true query values. Temperature and reasoning parameters are not overridden. Output limits are 24 000 tokens for Astra, 64 000 for Opus, and 65 536 for Flash and Argon.ssss

## S6. Size study details

For the separate $\varepsilon_{LUMO}$ size study, different 12500 molecules were sampled from QM9. From them, the samples of the size 125, 1250 and 12500 were randomly drawn (seed=2032025) and divided to the training, validation, and test sets in 9:1:2.5 proportion. These sets were then employed to train the models following the same protocol as in the main comparison, except for the learning rate and early stopping patience for fine-tuning and LoRA, which were set to $5\times10^{-6}$ and 6 for 1250 and 12500 points correrspondingly.

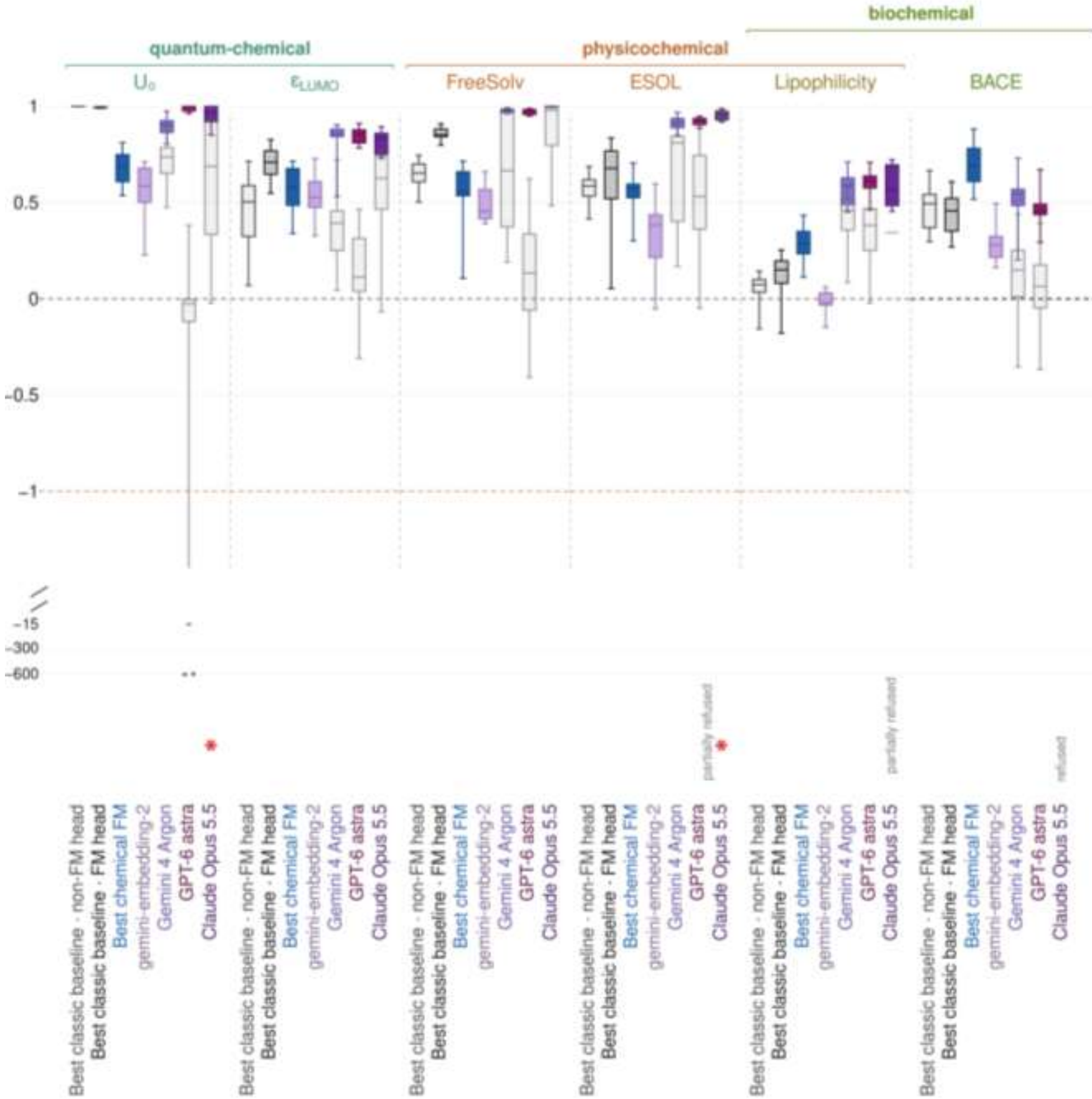

***Supplementary fig. 1 | Molecular property prediction from 60 labelled datapoints, shown over the full performance range****. Scores are $R^2$ for regression and $2 \times$ ROC-AUC $- 1$ for classification (BACE); dashed lines mark randomly guessing the class for classification or predicting the mean for regression. Boxes show the median and interquartile range; whiskers span the observed scores. Individual point markers are omitted. Grey boxes show performance after shuffling the labels of labelled datapoints. Red asterisks (∗) denote suspected memorization. 'Refused' indicates that the model declined to predict a target property in all fifteen attempts; 'Partially refused' indicates that the model declined to predict a target property in some attempts. Grey dots denote outliers.*

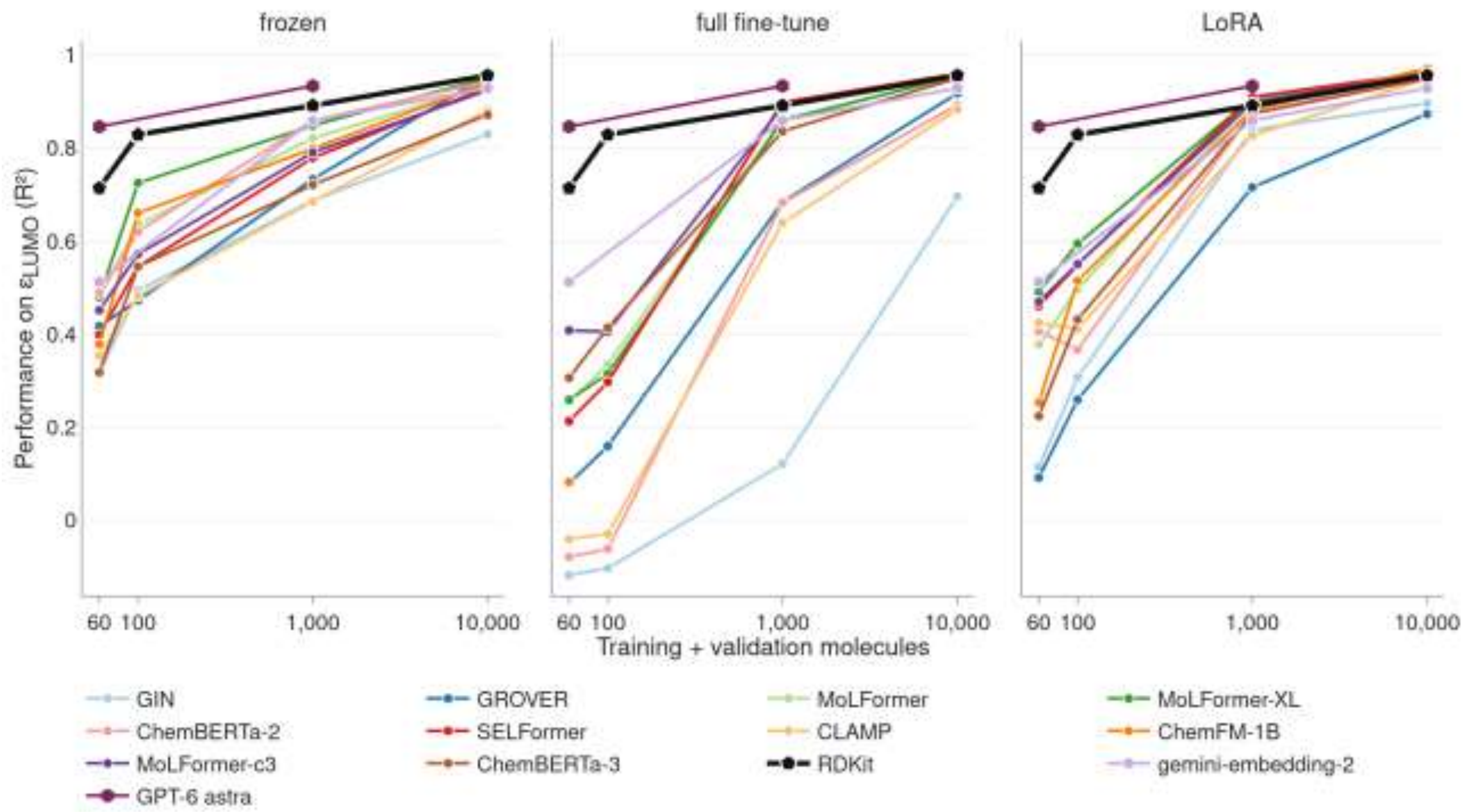

***Supplementary Fig. S2. Effect of labelled dataset size on $\varepsilon_{LUMO}$ prediction, shown over the full $R^2$ range.*** *Mean $R^2$ across five repeats is shown for frozen representations (left), full fine-tuning (middle) and low-rank adaptation (LoRA, right), at 60, 100, 1000 and 10000 molecules in training and validation subsets together. Black lines show RDKit descriptors with the best-performing prediction head. GPT-6 Astra's performance corresponds to few-shot prediction and is identical in all three plots. ChemFM-1B diverged to NaN in all our full-fine-tuning attempts, except for 60 labelled datapoints.*

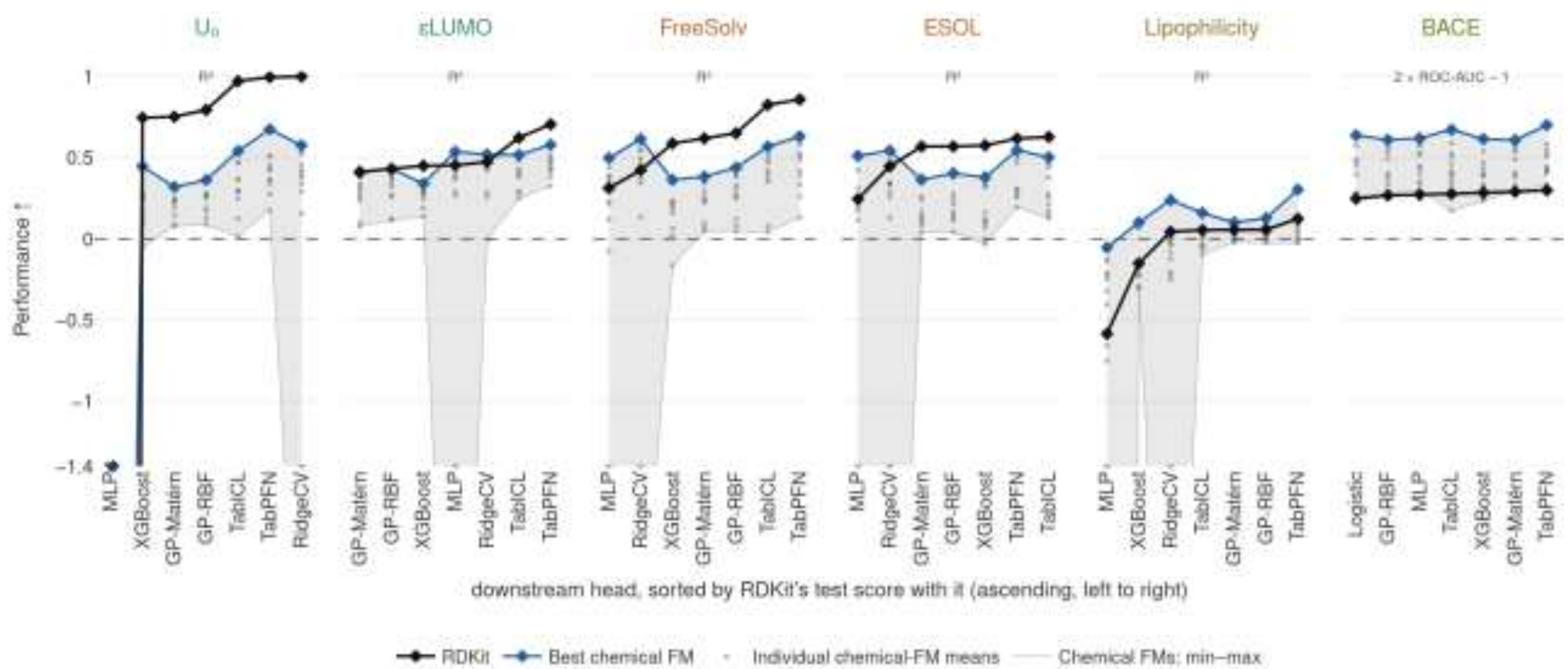

***Supplementary Fig. S3. Test performance of various heads for rdkit/chemical foundation models.*** *Black line and markers correspond to rdkit performance with the chosen head. Blue line and markers correspond to the chemical FM, showing the best test score with the corresponding head. Grey dots denote other chemical FMs. Points with performance below -1.4 are denoted by downward triangles.*